\documentclass[aps,
							 pre,
							 superscriptaddress, 
							 twocolumn,
               showpacs]{revtex4-1} 
\usepackage{graphicx, amsmath}

\newcommand\epsone{\epsilon_{I}}
\newcommand\epstwo{\epsilon_{II}}
\renewcommand\L{\mathcal{L}}
\newcommand\hamm{\mathcal{H}}
\newcommand\C{\mathcal{C}}

\newcommand\ER{Erd\H{o}s-R\'enyi\ }

\begin{document}

\title{Spatial network surrogates for disentangling complex system structure from spatial embedding of nodes}

\author{Marc Wiedermann}
\email{marcwie@pik-potsdam.de}
\affiliation{Potsdam Institute for Climate Impact Research --- P.O. Box 60 12
	03, 14412 Potsdam, Germany, EU}
\affiliation{Department of Physics, Humboldt University --- Newtonstr. 15,
	12489 Berlin, Germany, EU}
\author{Jonathan F. Donges}
\affiliation{Potsdam Institute for Climate
	Impact Research --- P.O. Box 60 12 03, 14412 Potsdam, Germany, EU}
\affiliation{Stockholm Resilience Centre, Stockholm University --- Kr\"aftriket 2B, 114 19 Stockholm, Sweden, EU} 
\author{J\"urgen Kurths}
\affiliation{Potsdam Institute for Climate Impact Research --- P.O. Box 60 12
	03, 14412 Potsdam, Germany, EU}
\affiliation{Department of Physics, Humboldt University --- Newtonstr. 15,
	12489 Berlin, Germany, EU}
\affiliation{Institute for Complex Systems and Mathematical Biology,
  University of Aberdeen --- Aberdeen AB24 3FX, UK, EU}
\affiliation{Department of Control Theory, Nizhny Novgorod State University ---
	Gagarin Avenue 23, 606950 Nizhny Novgorod, Russia}
\author{Reik V. Donner}
\affiliation{Potsdam Institute for Climate Impact Research --- P.O. Box 60 12
	03, 14412 Potsdam, Germany, EU}

\date{\today}

\begin{abstract}
Networks with nodes embedded in a metric space have gained increasing interest
in recent years. The effects of spatial embedding on the networks' structural
characteristics, however, are rarely taken into account when studying their
macroscopic properties. Here, we propose a hierarchy of null models to generate
random surrogates from a given spatially embedded network that can preserve
global and local statistics associated with the nodes' embedding in a metric
space. Comparing the original network's and the resulting surrogates' global
characteristics allows to quantify to what extent these characteristics
are already predetermined by the spatial embedding of the nodes and links. We
apply our framework to various real-world spatial networks and show that the
proposed models capture macroscopic properties of the networks under study much
better than standard random network models that do not account for the nodes'
spatial embedding. Depending on the actual performance of the proposed null
models, the networks are categorized into different classes. Since many
real-world complex networks are in fact spatial networks, the proposed approach
is relevant for disentangling underlying complex system structure from spatial
embedding of nodes in many fields, ranging from social systems over
infrastructure and neurophysiology to climatology.
\end{abstract}


\maketitle

\section{Introduction}
Many, if not most complex systems that exhibit a network structure are
spatially embedded in some metric space~\cite{barthelemy_spatial_2011}.
Examples of large interest include on the one hand structural networks, such as
social~\cite{capocci_preferential_2006}, transportation and distribution
\cite{banavar_size_1999,chan_urban_2011,jarvis_resource_2015}, communication
\cite{buldyrev_catastrophic_2010}, or electricity
networks~\cite{buzna_evolution_2009} with links corresponding to connections
between the entities represented by the networks' nodes. On the other hand,
functional networks with links indicating functional, mostly statistical,
interdependencies between the dynamics of individual nodes have been studied in the context of
functional brain~\cite{achard_resilient_2006,zhou_hierarchical_2006} or climate
networks~\cite{donges_complex_2009,donges_backbone_2009}.

A variety of network measures ranging from individual node's properties such as
degree and shortest-path betweenness to global characteristics such as
clustering coefficients and average path length are commonly utilized to
quantify the structural properties of a system under
study~\cite{newman_structure_2003,boccaletti_complex_2006}. Many studies aim to
classify the investigated networks into different categories, such as
small-world networks~\cite{watts_collective_1998} and subclasses
thereof~\cite{amaral_classes_2000,barabasi_scale-free_2009} by means of the
aforementioned topological characteristics. 

In fact, many of the complex systems commonly studied using network
theoretical methods are in fact spatial networks with nodes and links embedded
in some metric space~\cite{barthelemy_spatial_2011}, e.g., the Earth's surface
for infrastructure or climate networks~\cite{tsonis_what_2006}. Most studies,
however, do not take into account the possible influence of a network's spatial
structure on its resulting micro- or macroscopic characteristics. Thus, it often
remains unquantified whether a certain categorization of a network, such as the
small-world property, is to some extent already explicable as emerging from the
network's spatial embedding alone. Specifically, established random network
models that may be used to assess whether a network follows a certain rule of
construction solely preserve topological characteristics, such as the link
density in \textit{\ER random graphs}~\cite{erdos_evolution_1960} or the degree
sequence in the \textit{configuration model}~\cite{molloy_critical_1995}. 

To classify possible types of spatially embedded networks, several models have
been proposed that generate random surrogates from a given spatially embedded
node sequence by, e.g., randomly distributing links according to the spatial
distance between
nodes~\cite{barnett_spatially_2007,heitzig_node-weighted_2012}, setting a
prescribed linkage probability between nodes to address boundary effects in
climate networks~\cite{rheinwalt_boundary_2012} or optimizing the
length-dependent costs related to the construction of a link in power
grids~\cite{schultz_random_2014}. A variety of studies introduced random
network models to specifically reproduce statistics associated with brain
networks. Most of these models are of generative nature and set up artificial
networks that are then compared with observations made from data. Some models
are designed such that connections between regions of similar input are favored
but long-range connections are penalized~\cite{vertes_simple_2012}, e.g., by
optimizing the interplay between range-dependent wiring cost and processing
efficiency~\cite{chen_trade-off_2013}. In other models the connectivity between
nodes or areas of the brain depends on gravitational forces between these
different areas~\cite{song_spatial_2014} and even other models are designed
such that links are put between nodes depending on their participation
coefficient~\cite{rubinov_wiring_2015}. 

All of the above mentioned models, however, have been primarily designed to assess
and reproduce construction principles behind certain types of complex networks
and their underlying mechanism are usually tailored to a specific application.

In order to explicitly study the general influence of a network's spatial
embedding on its resulting macroscopic characteristics we propose here a set of
random network models to create surrogates that preserve certain geographical
and topological features of these given networks. The surrogates are
constructed by iteratively rewiring the original network while preserving a set
of its geographical features. In particular, one model, which will be called
\textit{GeoModel I} hereafter, preserves, in addition to the degree sequence,
the global link-length distribution. A second model referred to as
\textit{GeoModel II} additionally preserves for each node the length
distribution of the links connected to it and, hence, imposes an even stronger
spatial constraint on the rewiring process. The resulting surrogate networks
allow for evaluating to what extent observed macroscopic properties of a given
network are explicable by geometric constraints inflicted on the network's
structure, while no assumptions on the specific construction principles are
necessary.

We apply our method to a number of real-world complex networks: the US airline
network, the US interstate network, the Internet~\cite{gastner_spatial_2006},
the Scandinavian power grid~\cite{menck_how_2014}, a world trade
network~\cite{lenzen_mapping_2012}, and the road network of a German city.  
Additionally, we study the application of our models to a random geometric
graph~\cite{penrose_random_2003} and an \ER
network~\cite{erdos_evolution_1960}. For comparison, we construct iteratively
rewired surrogate networks that only preserve topological characteristics of
the given networks, namely the mean degree on the one hand and the degree
distribution on the other hand.

Our study reveals that the macroscopic properties of a certain set of networks
are only reproduced by applying either of the two geometrically constrained
models proposed in this work, while the consideration of topological features
alone is not sufficient.  Generally, preserving the global link length
distribution and, hence, applying GeoModel I already reproduces well the average
path length of a given network. In order to additionally reproduce the global
clustering coefficient, the per-node link length distributions also need to be
taken into account and, hence, the application of GeoModel II becomes necessary.

The remainder of this paper is organized as follows. Section~\ref{sec:methods}
introduces the algorithms behind the random network models proposed in this
work as well as the network characteristics that are used to evaluate their
performances.  Section~\ref{sec:data} gives an overview on the network data
that is investigated and Sec.~\ref{sec:results} showcases the results of the
study. Section~\ref{sec:conclusion} presents our conclusions and an outlook on
future directions of research.

\section{Methods}\label{sec:methods}
\subsection{Preliminaries}
Consider a network $G=(V, E)$ with given sets of nodes ($V$) and links ($E$).
Each node is labeled with a natural number $i=1,2,\ldots,N$, with $N$ being the
total number of nodes in the network. The network is represented by its
adjacency matrix $\mathbf{A}$ with entries $A_{ij}=1 \text{if}\ \{i,j\}\in E$,
and $A_{ij}=0$ otherwise. Thus, we study here the case of undirected and
unweighted networks without self-loops and multiple links between nodes.
Additionally, each node is assigned a position $\mathbf{x}_i$ in some metric
space of dimension $d$. In the applications presented in this work, nodes are
either embedded on the surface of a sphere, i.e., the Earth's surface, or in a
Cartesian coordinate system. In the first case, the position of a node is
determined by its latitudinal and longitudinal coordinates $\lambda_i$ and
$\phi_i$ and, hence, $\mathbf{x}_i=(\lambda_i, \phi_i)$. In the second case, a
node's position is given by its Cartesian coordinates $x_i$ and $y_i$ with
respect to some arbitrarily chosen origin, $\mathbf{x} = (x_i, y_i)$.  The
$N\times N$ distance matrix $\mathbf D$ then gives the distance between all
nodes in the network. For the case of a spherical coordinate system, its
entries $d_{ij}$ are computed as the great circle distances between nodes, 
\begin{align}
	d_{ij} = R
	\arccos(\sin\lambda_i\sin\lambda_j+\cos\lambda_i\cos\lambda_j\cos\Delta_{ij})
\end{align}
with $\Delta_{ij} = \phi_j - \phi_i$.
$R$ denotes the radius of the sphere, which is rescaled to unit length in
all applications and, hence, we set $R=1$.
For the case of a Cartesian coordinate system, the entries of $\mathbf{D}$ are given by the
Euclidean distance between two nodes,
\begin{align}
	d_{ij} = \sqrt{(x_j - x_i)^2+ (y_j - y_i)^2}.
\end{align}
From this the local cumulative distribution function $P_i(l)$ of link lengths
$l$ of node $i$ follows directly as
\begin{align}
	P_i(l) = \frac{1}{k_i} \sum_{j} A_{ij} \Theta(l-d_{ij}).
\end{align}
Here, $k_i=\sum_j A_{ij}$ is the degree, i.e., the number of neighbors, of a
node $i$ and $\Theta(\cdot)$ is the Heaviside function. The global cumulative link length distribution $P(l)$ in the network is then given as
\begin{align}
	P(l) = \frac{1}{2M} \sum_i k_i P_i(l),
\end{align}
with $M=|E|$ denoting the total number of links in the network.

\subsection{Complex network characteristics}
To characterize the macroscopic structure of the networks under study as well
as their corresponding random surrogates, we rely on two commonly used global
network measures, the global clustering coefficient $\C$ and the average path
length $\L$~\cite{newman_structure_2003}. An evaluation of both
measures is commonly used to classify a network under study as, e.g., a
small-world network, which is defined to display a high clustering coefficient while at the same time showing a low average path
length~\cite{watts_collective_1998}.

\subparagraph{Global clustering coefficient.}
The global clustering coefficient $\C$ gives the probability to find
connected triples, i.e., closed triangles formed by links in the
network adjacent to a randomly selected node~\cite{watts_collective_1998}. It is defined as the
arithmetic mean taken over all local clustering coefficients,
\begin{align}
	\C = \frac{1}{N} \sum_i \C_i
\end{align}
with 
\begin{align}
	\C_i = \frac{1}{k_i(k_i-1)} \sum_{\substack{ j\neq k\\ j,k\neq i}} A_{ij}A_{jk}A_{ki}.
\end{align}
Note that $\C_i$ is only defined if $k_i > 1$. Otherwise we
set $\C_i=0$.

\subparagraph{Average path length.}
The average path length $\mathcal L$ gives the average number of edges along shortest paths
between two randomly chosen nodes. Given that $\mathcal L_{ij}$ denotes the number of
such steps between two nodes $i$ and $j$, the average path length follows as
\begin{align}
	\mathcal L = \frac{1}{N(N-1)}\sum_{i\neq j} \mathcal L_{ij}.
\end{align}
In the case when there exists no path between $i$ and $j$ we set $\mathcal L_{ij} =
N-1$.
\subparagraph{Hamming distance.}
Consider two undirected networks $G=(V, E)$ and $G'=(V, E')$ with a common set of nodes and
the same number of links $M=|E|=|E'|$ represented by adjacency matrices $\mathbf{A}$ and
$\mathbf{A}'$. The Hamming distance $\hamm$ then provides a measure of the
dissimilarity between the two sets of links $E$ and
$E'$~\cite{radebach_disentangling_2013,hamming_error_1950},
\begin{align}
	\hamm = \frac{1}{4M} \sum_{i,j} |A_{ij}'-A_{ij}| \in [0,1].
\end{align}
A Hamming distance of $\hamm = 0$ implies that the two networks $G$ and $G'$
have identical sets of links ($E=E'$), while $\hamm=1$ indicates that the
two sets of links have entirely dissimilar entries ($E\cap E'=\emptyset$). In
the scope of this work, $\hamm$ is utilized to assess the
dissimilarity between a network under study and the surrogate networks that are
created from it. Hence, we aim to maximize the Hamming distance between a
network and its surrogates while at the same time evaluating the degree of
similarity between the global clustering coefficients and average path lengths of
the original and the random networks.

We note that the network characteristics introduced above form only a small
subset of possible measures that could be investigated in the course of this
work. Among others, link-weighted quantities such as the weighted average
shortest path length~\cite{latora_economic_2003} or the weighted clustering
coefficient~\cite{newman_scientific_2001} (of which there exist a variety of
definitions~\cite{saramaki_generalizations_2007}) have become of increasing
interest when quantifying the topology of real-world complex
networks~\cite{barrat_architecture_2004}. However, as we aim here to
inter-compare the resulting quantities among the different networks under
study, we would have to define a concise weighting scheme that is applicable
simultaneously to all of them.  Note that, for example, in airline networks
links are commonly weighted by the number of available seats or the number of
flights on a route between two
airports~\cite{bagler_analysis_2008,barrat_weighted_2004}.  On the other hand
road networks may be weighted by, e.g., the total traffic flow along certain
roads~\cite{de2007structure} while trade networks are weighted by the total
trade volume between two partners~\cite{fagiolo_topological_2008}. Thus,
macroscopic quantities computed from different networks weighted according to
these definitions would not be straightforward to compare as they describe
different physical entities.  Even neglecting these usually employed
definitions of link weights, a consistent weighting that is purely based on
distances is similarly hard to perform. While infrastructure networks (such as
airline networks) should be weighted linearly according to the spatial distance
between nodes, link weights in communication networks (such as the Internet)
should correspond to the inverse distance, i.e., the per-link
efficiency~\cite{latora_economic_2003}.  For networks such as the world trade
network or synthetic networks as the Erd\H{o}s-R\'enyi graph and the random
geometric graph a weighting based on node-distances would further lack a proper
and concise interpretation. 

In the course of this work we thus restrict ourselves to the assessment of well
established unweighted network measures which were already successfully applied
to quantify and classify the topology of spatially embedded
networks~\cite{crucitti_centrality_2006,gastner_spatial_2006}. This procedure
provides results that are easily comparable and interpretable among the
different networks studied in this work. A refinement of this study by well
defined weighted network characteristics remains as an important
subject of future research.

\subsection{Random network models}
\begin{figure}[t]
	\includegraphics[width=.9\linewidth]{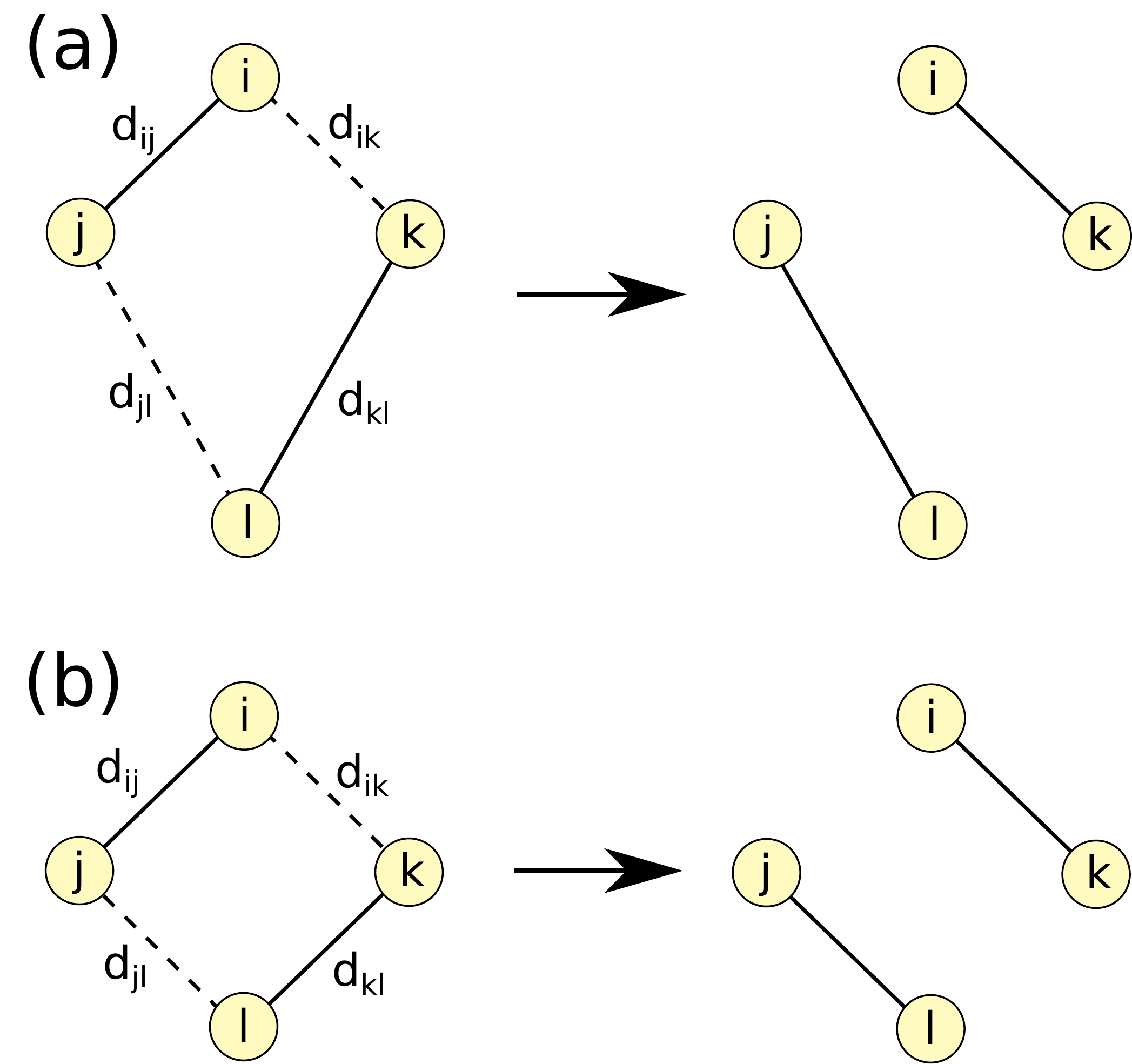}
	\caption{(Color online) Sketch of the rewiring process that generates
		randomized surrogates of a given original network by applying GeoModel I (a)
		and GeoModel II (b). Nodes $i$, $j$, $k$ and
		$l$ are drawn at random, such that $i$ is linked to $j$ and $k$ is linked
		to $l$ (solid lines) but no links are present between $i$ and $k$,
		and $j$ and $l$ (dashed lines). According to the chosen network model the distances
		$d_{\bullet\bullet}$ between the nodes are evaluated. If the nodes form
		an approximate kite (a) or diamond (b) with the connections between them present as depicted,
		previous links are replaced by 
		links connecting $i$ with $k$ and $j$ with $l$. 
	 	\label{fig:fig_01}}
\end{figure}
We generate random network surrogates from a given real-world network by
applying four different algorithms. Two of them, random link switching and
random rewiring do not take into account any spatial embedding of the
network's nodes, whereas this consideration is an explicit part of the two novel
models, GeoModel I and GeoModel II.

The general structure of the algorithms is described as follows. Starting from a
copy $\mathbf A'$ of the original network's adjacency matrix $\mathbf A$:
\begin{itemize}
	\item[(i)] Draw four distinct nodes $i$, $j$, $k$, $l$ uniformly at random
		from $V$.
	\item[(ii)] Depending on the applied random network model under study, check
		whether a certain condition $\mathbf C$ is TRUE. If $\mathbf C$ is
		FALSE return to step (i).
	\item[(iii a)] (applies to random rewiring) If $\mathbf C$ is TRUE, break the link connecting $i$
		with $j$ and establish a new link connecting $k$ with $l$. Hence, 
		$A_{ij}'=A_{ji}'=1\rightarrow 0$ and $A_{kl}'=A_{lk}'=0\rightarrow 1$.
	\item[(iii b)] (applies to all other random network models) If $\mathbf C$ is TRUE, break the links connecting $i$
		with $j$ and $k$ with $l$ and establish links connecting $i$ with $k$ and
		$j$ with $l$.
		Hence, 
		$A_{ij}'=A_{ji}'=1\rightarrow 0$, $A_{kl}'=A_{lk}'=1\rightarrow 0$,
		$A_{ik}'=A_{ki}'=0\rightarrow1$,
		and $A_{jl}'=A_{lj}'=0\rightarrow1$.
	\item[(iv)] As long as a certain number of rewirings $r$ is not reached return
		to (i) with the modified adjacency matrix $\mathbf A'$. 
\end{itemize}
The resulting modified copy $\textbf{A}'$ of the original network's adjacency
matrix $\textbf{A}$ is then returned for further evaluation. In the following,
we introduce the explicit form of the conditions $\textbf C$ for a rewiring
process to take place.

\subsubsection{Random Rewiring}
Random rewiring, the simplest case, takes place if a link exists between the
randomly drawn nodes $i$ and $j$, but no link exists between $k$ and $l$.
Hence,
\begin{align}
	\textbf C=A_{ij}' \land\lnot A_{kl}'.
\end{align}
Here, $\land$ denotes the truth-functional operator of logical
conjunction and $\lnot$ denotes the logical complement, i.e., $\lnot1=0$ and
$\lnot0=1$. Note that in the scope of this work the adjacency matrices' entries
$A_{ij}=1$ are interpreted as the logical 1 (TRUE) and entries $A_{ij}=0$ are
interpreted as the logical 0 (FALSE). Thus, the resulting value of $\textbf C$
is also a logical value being either TRUE or FALSE.

The definition of $\textbf C$ is then plugged into step (ii) and (iii a) in the
above algorithm and depending on its value a new set of four nodes is drawn or
a rewiring process takes place. Random rewiring solely preserves the average
degree $K=N^{-1}\sum_i k_i$ in the network and, hence, after sufficiently many
rewiring steps, the resulting surrogate network converges to an \ER random
graph~\cite{erdos_evolution_1960}.

\subsubsection{Random Link Switching}
In addition to the mean degree $K$, random link switching preserves the local
degree of each node in the network as well, but still neglects any aspect of a
network's spatial
embedding~\cite{zamora-lopez_reciprocity_2008,artzy-randrup_generating_2005}.
This framework has in the past already been successfully applied to study
topological properties of real-world systems such as the
Internet~\cite{maslov_detection_2004} or protein
networks~\cite{maslov_specificity_2002}. Here, for the four nodes drawn in step
(i) of the construction algorithm, we need to ensure that $i$ is linked with
$j$ and $k$ is linked with $l$, but $i$ and $k$ as well as $j$ and $l$ are not
yet connected (Fig.~\ref{fig:fig_01}). Hence, the condition $\textbf C$ reads 
\begin{align}
\textbf C = \textbf C_1 = A_{ij}' \land A_{kl}' \land \lnot A_{ik}'
\land\lnot A_{jl}'.
\end{align}
As the degree of each node is preserved, the resulting
surrogate networks relate to the results one would obtain from the configuration
model~\cite{molloy_critical_1995}. However, for the present case the surrogate
networks display no self-loops or multiple links between nodes.

\subsubsection{GeoModel I}
\begin{table*}[t]
\centering
\setlength{\tabcolsep}{.1cm}
\begin{tabular}{lrrrrrrrrr}
	Name & \multicolumn{1}{c}{$N$} & \multicolumn{1}{c}{$M$}
	&\multicolumn{1}{c}{$K$}& \multicolumn{1}{c}{$\rho$} &
	\multicolumn{1}{c}{$\mathcal{C}$} & \multicolumn{1}{c}{$\mathcal{L}$} &
	\multicolumn{1}{c}{$\epsone$} & \multicolumn{1}{c}{$\epstwo$} & Grid Type\\\hline
US airline & 190 & 837 & 8.86 & 0.0466 & 0.679 & 2.176 & 0.04 & 0.07 & Spherical\\
Internet & 13.372 & 28.253 & 4.23 & 0.0003 & 0.423 & 3.630 & 0.04 & 0.01 & Spherical \\
US interstate & 935 & 1.315 & 2.82 & 0.0030 & 0.107 & 20.207 & 0.17 & 0.24 & Spherical \\
Scandinavian power grid & 236 & 318 & 2.71 & 0.0115 & 0.084 & 9.156 & 0.16 & 0.27 & Cartesian \\
World trade & 186 & 7.043 & 76.14 & 0.4094 & 0.815 & 1.594 & 0.02 & 0.04 & Spherical \\
Urban roads (Eschwege) & 855 & 1.174 & 2.75 & 0.0032 & 0.050 & 18.313 & 0.15 & 0.22 & Spherical \\
Random geometric graph & 2.000 & 5.493 & 5.50 & 0.0027 & 0.588 & 30.428 & 0.11 &
0.13 & Cartesian \\
Erd\H{o}s-R\'enyi graph & 2.000 & 5.493 & 5.50 & 0.0027 & 0.003 & 4.643 & 0.01 &
0.01 & Cartesian \\\hline
\end{tabular}
\caption{Overview of all networks investigated in this study including their
	number of nodes $N$ and links $M$, average degree $K$, link density $\rho$, global clustering coefficient $\mathcal{C}$,
	and average path length $\mathcal{L}$. $\epsone$ and $\epstwo$ denote the relative
	tolerances that are chosen for generating random network surrogates from
	GeoModel I and GeoModel II, respectively (see text). \label{tab:tab_01}}
\end{table*}
In addition to the above criterion $\textbf C_1$, GeoModel I aims to also preserve the global
link length distribution $P(l)$. Hence, the potentially newly established links must
be of the same length as those that are removed from the network. This means
that the four randomly drawn nodes $i$, $j$, $k$ and $l$ must form a kite with
exactly one link present at each of the two sides of the same length
(Fig.~\ref{fig:fig_01}(a)). Since the
nodes are usually embedded in a continuous domain this equality can only be
fulfilled up to a certain accuracy. We hence demand that the newly established links
have approximately the same length as the existing ones with some tolerance
$\epsilon$. In other words, the difference in lengths between the present and
potentially established links should not exceed a certain fraction $\epsilon$
of the existing links' lengths. 
Thus, the following condition must be fulfilled,
\begin{align}
	\textbf C&=\textbf C_1 \land \textbf C_2\ \mbox{with} \label{eqn:geoone_cond}\\
	\textbf C_2&=\Theta(\epsilon d_{ij}- |d_{ij}-d_{ik}|)\land \Theta(\epsilon d_{kl}
	-|d_{kl}-d_{jl} |).\label{eqn:geoone_firstcond}
\end{align}
Here, $\epsilon$ measures the maximum allowed relative deviation in length
between the existent and potentially newly established links. As for the
entries of the adjacency matrices $\textbf{A}$, a value of $\Theta(\cdot)=1$
here denotes the logical 1 (TRUE) and $\Theta(\cdot)=0$ correspondingly
represents the logical 0 (FALSE).

GeoModel I preserves the degree distribution in the same way as random link
switching, but in addition approximately preserves the global link length
distribution $P(l)$. The ensemble $\Omega_{GMI}$ of possible surrogates
constructed by GeoModel I therefore forms a subset of the ensemble
$\Omega_{rls}$ of all those surrogates possibly constructed from random link
switching, $\Omega_{GMI}\subseteq\Omega_{rls}$.  Generally, it is to be
expected that with an increasing $\epsilon$ the Hamming distance $\hamm$
between the original networks and its surrogates increases.  However, an
increase in $\epsilon$ also induces larger deviations between the link length
distributions of the original and surrogate networks. Hence, we aim to estimate
the maximum meaningful value of $\epsilon$ by using a Kolmogorov-Smirnoff (KS)
test~\cite{press_numerical_1996} (see Appendix~\ref{sec:appendix} for
details), demanding that for an ensemble of $n$
surrogate networks the resulting link length distributions are statistically
indistinguishable from that of the original network in $95\%$ of all cases
under a confidence level of $\alpha=95\%$.  

\subsubsection{GeoModel II}
In order to not only preserve the global but also the local (per-node)
link length distributions $P_i(l)$, we demand that the two links to be removed and the
two links to be established all have approximately the same length. Hence,
the nodes $i$, $j$, $k$ and $l$ form a diamond. That way, none of the lengths of
links emerging from either of the four nodes is changed under each rewiring
step. As above, in most situations this criterion can only be fulfilled approximately and we utilize,
for convenience, the same parameter $\epsilon$ to extend the previous conditions
$\textbf C_1$ and $\textbf C_2$ as
\begin{align}
	\textbf C&=\textbf C_1 \land \textbf C_2 \land \textbf C_3\ \mbox{with}\label{eqn:geotwo_cond}\\
	\textbf C_3&=\Theta(\epsilon\max(d_{ik},
	d_{jl})-|d_{ik}-d_{jl}|)\label{eqn:geotwo_firstcond}.
\end{align}
Thus, the difference in length of the newly established links (and therefore
also the difference in lengths of the existing links) must not be larger than a
certain fraction $\epsilon$ of their respective maximum length. For our studies we decided to
take the maximum of $d_{ik}$ and $d_{jl}$ as the reference scale of the tolerance
window. However, other choices, such as the minimum value or the arithmetic
mean of the two, might also be possible and would result in different optimal
values of the tolerance parameter $\epsilon$. A detailed investigation on the
effect of the actual definition of the link length that is chosen as a reference
remains as a subject of future research.

Again, the ensemble $\Omega_{GMII}$ of all possible surrogates constructed from
GeoModel II forms a subset of all possible surrogates constructed from GeoModel I and
random link switching, since it only imposes a further condition in addition
to the already employed ones. Hence,
$\Omega_{GMII}\subseteq\Omega_{GMI}\subseteq\Omega_{rls}$.

\section{Data}\label{sec:data}
We consider different real-world networks to illustrate the performance of our
algorithms and test to what extent macroscopic characteristics are recaptured
by random network surrogates that take into account spatial constraints on the
distribution of links in the network. We first investigate three different
previously studied infrastructure networks~\cite{gastner_spatial_2006}: the US
airline network with nodes displaying airports and links indicating flights
scheduled between them, the US interstate network with links representing
highways and nodes serving as country borders, termination points and
intersections between highways, and the Internet with nodes corresponding to
autonomous systems around the globe where links stand for data connections
between them. Contrasting the case of the interstate network, we also study an
infrastructure network of smaller spatial scale by retrieving the urban road
network of a German small-town (Eschwege) from \url{www.openstreetmap.org}
(assessed 2012-01-30). Here, nodes again represent intersections and links are
roads connecting them.  Moreover, we apply our framework to the Scandinavian
power grid, where links represent high voltage transmission lines and nodes are
transformation stations or power plants~\cite{menck_how_2014}. The latter two types of
networks have been intensively studied in the framework of complex network
theory and the understanding of their global properties has been reported as
crucial since these strongly determine their local behavior, e.g., the
robustness to failures of single
nodes~\cite{albert_structural_2004,crucitti_topological_2004,kinney_modeling_2005}.
Finally, we study the world trade network of 2009 with nodes representing the
center of a country and links indicating trade between
them~\cite{lenzen_mapping_2012} as a representative of a non-physical, yet
spatially embedded transaction network.

For comparison with these real-world networks, we also study synthetic networks
with known properties, which serve as a benchmark for our analysis.
Particularly, we consider a random geometric graph with nodes put randomly on a
plain unit square~\cite{donges_analytical_2012,herrmann_connectivity_2003}. All
nodes with a spatial distance of less than 0.03 are connected to yield a
manageable density of links. We expect that this network's macroscopic
properties are only explainable by considering random network models that take
into account the spatial embedding of the nodes. For the sake of comparison, we
construct one realization of an \ER random graph with the same number and
position of nodes and the same number of links randomly put between them as in
the random geometric graph. As links are put without any relation to spatial
distances, the simplest network model, i.e., random rewiring, should already
capture this network's macroscopic features.  

A summary of all networks included in this study together with each network's
number of nodes $N$ and links $M$ as well as further network parameters is presented in Tab.~\ref{tab:tab_01}.

\section{Results}\label{sec:results}
\begin{figure}[t]
	\includegraphics[width=.75\linewidth]{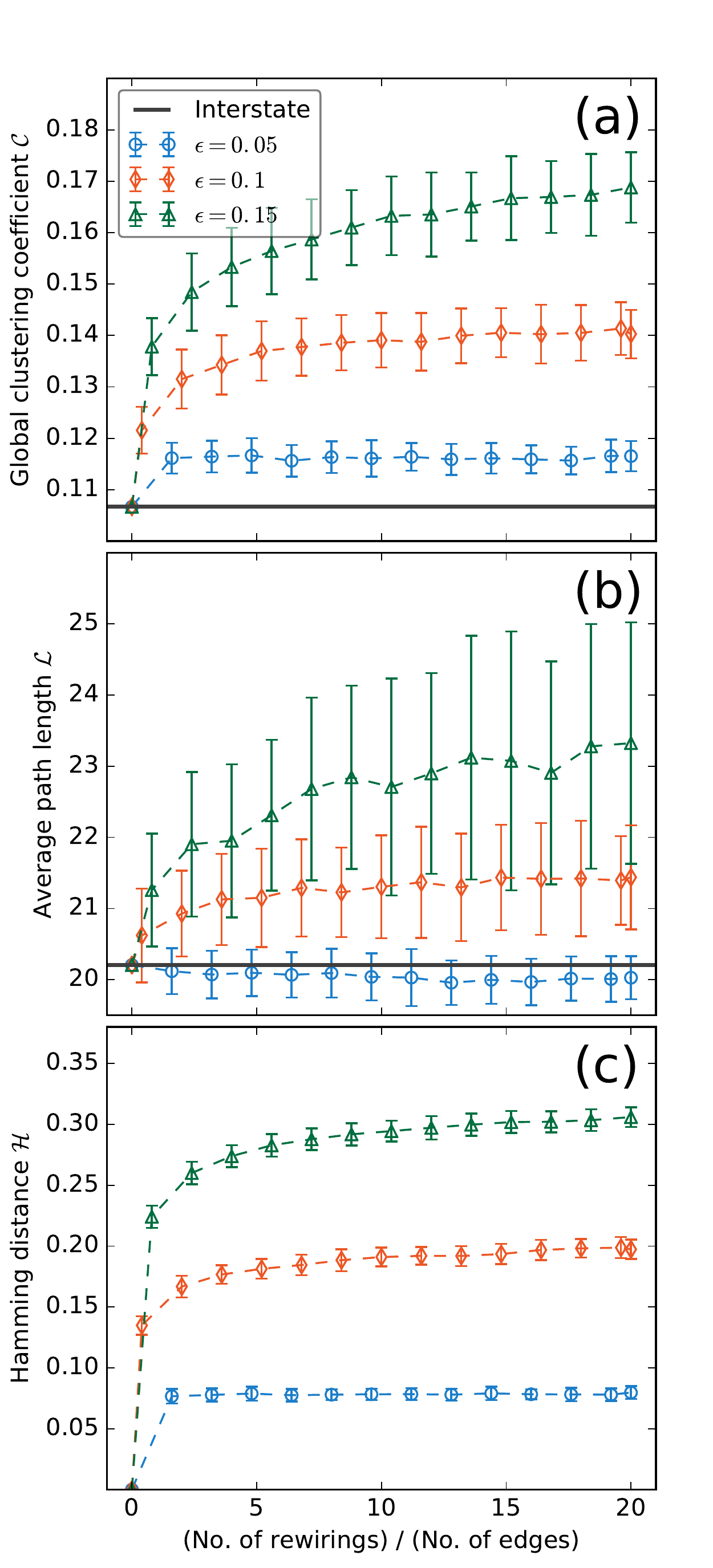}
	\caption{(Color online) Evolution of the global clustering coefficient $\C$ (a) and average
		path length $\L$ (b) with the number of rewirings for an ensemble of
		$n=100$ surrogates generated from the interstate network by applying
		GeoModel I and using different tolerances $\epsone$. Solid lines indicate the
		respective value of $\C$ and $\L$ of the interstate network itself. (c) Hamming distance $\hamm$ between the surrogate networks and the original
		network. Scatter symbols denote the mean value and error bars indicate one standard
		deviation of each measure.}\label{fig:fig_02}
\end{figure}

We now apply the four random network models introduced above to the different
real-world and synthetic networks under study. In a first step we illustrate
how to estimate a proper tolerance parameter $\epsilon$ for GeoModel I and
GeoModel II\@. Specifically, we illustrate the procedure for the example of the
US interstate network and the application of GeoModel I. We then discuss in
detail the results of all four network models applied to the US interstate and
the airline network and show to what extent macroscopic network characteristics
are reproduced by the different network models.  Finally, we present a
comprehensive intercomparison between all networks investigated in this study
by applying the different models to each real-world network. We evaluate, which
macroscopic features of a network can be reproduced by which model
and assign the real-world networks to different
classes, i.e., those for which spatial embedding plays a minor role when
estimating macroscopic properties and those where the spatial structure
explicitly needs to be taken into account.

For all cases discussed from now on, we construct an ensemble of $n=100$
surrogate networks for each network under study and iteratively rewire each
random model for $r=20M$ steps.

\subsection{Selection of the tolerance parameter $\epsilon$}
\begin{figure}[t]
	\includegraphics[width=.75\linewidth]{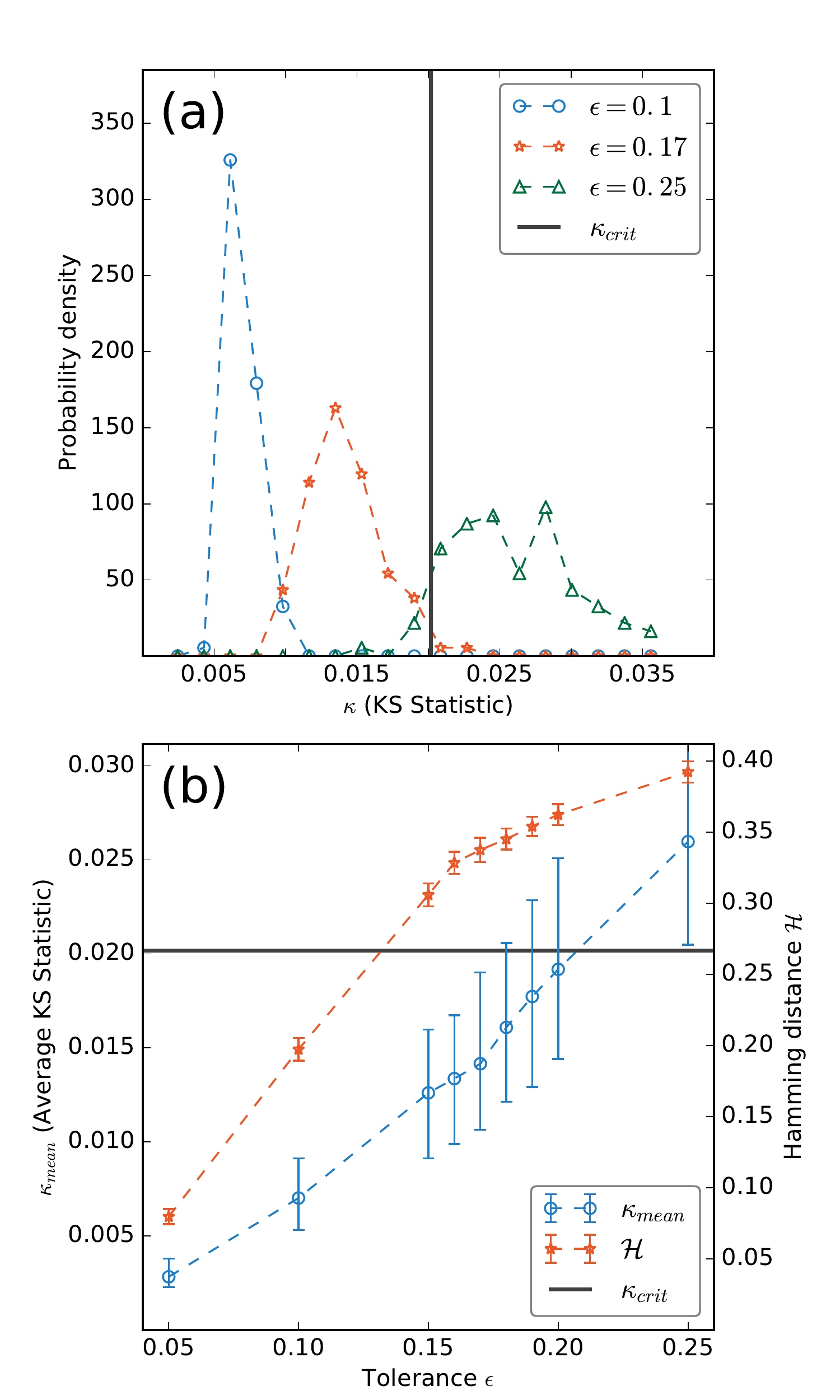}
	\caption{(Color online) (a) Distribution of KS statistics $\kappa$ measuring the
		maximum distance in the global cumulative link length distribution $P(l)$ between the
		interstate network and each of the $n=100$ network surrogates obtained by
		applying GeoModel I under different tolerances $\epsilon$ and $20M$ rewirings.
		(b) Average KS statistics $\kappa_{mean}$ and Hamming distance $\hamm$ after $20M$
		successful rewirings depending on the choice of tolerance $\epsilon$. Error
		bars denote one standard deviation for the Hamming distance and the 5th and
		95th
		percentile of the distribution of KS statistics. The solid line indicates
		the critical value $\kappa_{crit}$ below which the surrogates' and the original
		network's link length distributions are considered statistically
		indistinguishable under a confidence
		level
		of $\alpha=0.95$.}\label{fig:fig_03}
\end{figure}
The only free parameter in GeoModel I and GeoModel II is the tolerance parameter
$\epsilon$ in Eqs.~\eqref{eqn:geoone_firstcond} and
\eqref{eqn:geotwo_firstcond}, which determines
which link lengths are treated as being sufficiently similar. To illustrate the influence of
$\epsilon$ on our results, we apply GeoModel I to the US
interstate network and create $n=100$ surrogate networks that display the same degree
sequence and approximately the same global link length distribution $P(l)$ as the
original network. Figure~\ref{fig:fig_02} shows the typical evolution of
global clustering coefficient $\C$, average path length $\L$ and Hamming
distance $\hamm$ for different choices of $\epsilon$. As expected, we note that
for the lowest choice of $\epsilon$ ($\epsilon=0.05$) the surrogate networks' $\C$ and $\L$ are
closest to the values for the original network (Fig.~\ref{fig:fig_02}(a),(b)). However, in
that case, the Hamming distance displays low values around 0.075, meaning that
92.5\% of links in the original network are also present in the
surrogate networks (Fig.~\ref{fig:fig_02}(c)). With increasing $\epsilon$ the
values of $\hamm$ also increase and, hence, the surrogate networks become
increasingly dissimilar from the original network. At the same time, $\C$ and
$\L$ also differ more from their target values (Fig.~\ref{fig:fig_02}(a),(b)).

As GeoModel I aims to approximately preserve the global link length distribution
$P(l)$, we examine also the distribution of the KS statistics $\kappa$ for the
ensemble of surrogate networks at different tolerance parameters $\epsilon$
(Fig.~\ref{fig:fig_03}(a)). For low values of $\epsilon$, all cumulative link
length distributions are statistically indistinguishable with 95\% confidence, which results in values
of $\kappa$ being smaller than the critical value $\kappa_{crit}$. This value
indicates the upper bound of the confidence interval (Fig.~\ref{fig:fig_03}(a))
and is determined as the largest possible value that satisfies
Eq.~\eqref{eqn:k_crit}.  However, as already discussed above, the Hamming
distance $\hamm$ becomes very low for low $\epsilon$ and only a few links differ
between the original and the surrogate networks (Fig.~\ref{fig:fig_03}(b)). On
the other hand, for large $\epsilon$ most link length distributions are
dissimilar under the desired confidence level and, hence, the purpose of
GeoModel I is not fulfilled. We find that for $\epsilon=0.17$, $95\%$ of all
distributions are statistically equivalent with 95\% confidence and, hence,
GeoModel I
achieves its highest possible Hamming distance (Fig.~\ref{fig:fig_03}).

Following the same procedure, the optimal tolerance can be obtained for
GeoModel II
as well as for all other networks under study. It is important to note that the
values of $\epsilon$ generally differ between the two random network models as
for GeoModel II the additional criterion $\textbf C_3$ must
be fulfilled. We therefore denote $\epsone$ the optimal tolerance for GeoModel
I and
$\epstwo$ the respective optimal tolerance for GeoModel II.  A summary of all
tolerances for each network and random network model is given in
Tab.~\ref{tab:tab_01}. We note that the obtained values of $\epsilon$ differ
between $\epsilon=\epstwo=0.01$ for the Internet and $\epsilon=\epstwo=0.27$
for the power grid. Moreover, in most cases we find that $\epsone < \epstwo$.

The heterogeneity in the distribution of links in the original network seems to
seems to play a crucial role for the resulting value of $\epsilon$. The
interplay between the tolerance parameter and the shape of the cumulative
distribution function of link lengths as well as the number of nodes $N$ and
links $M$ should be addressed in future research. Further, we
note that the link length distributions of the networks under study are not
necessarily symmetric. Thus, more advanced statistical tests such as Anderson-Darling or Shapiro-Wilk
tests with potentially larger power than the Kolmogorov-Smirnoff test might improve the assessment of the statistical equivalence between the
surrogates' and the original networks' distributions~\cite{razali_power_2011}.
An assessment of these quantifiers is, however, beyond the scope of this work
and remains as a subject of future studies.

\subsection{Interstate network}
\begin{figure}[t]
	\includegraphics[width=.8\linewidth]{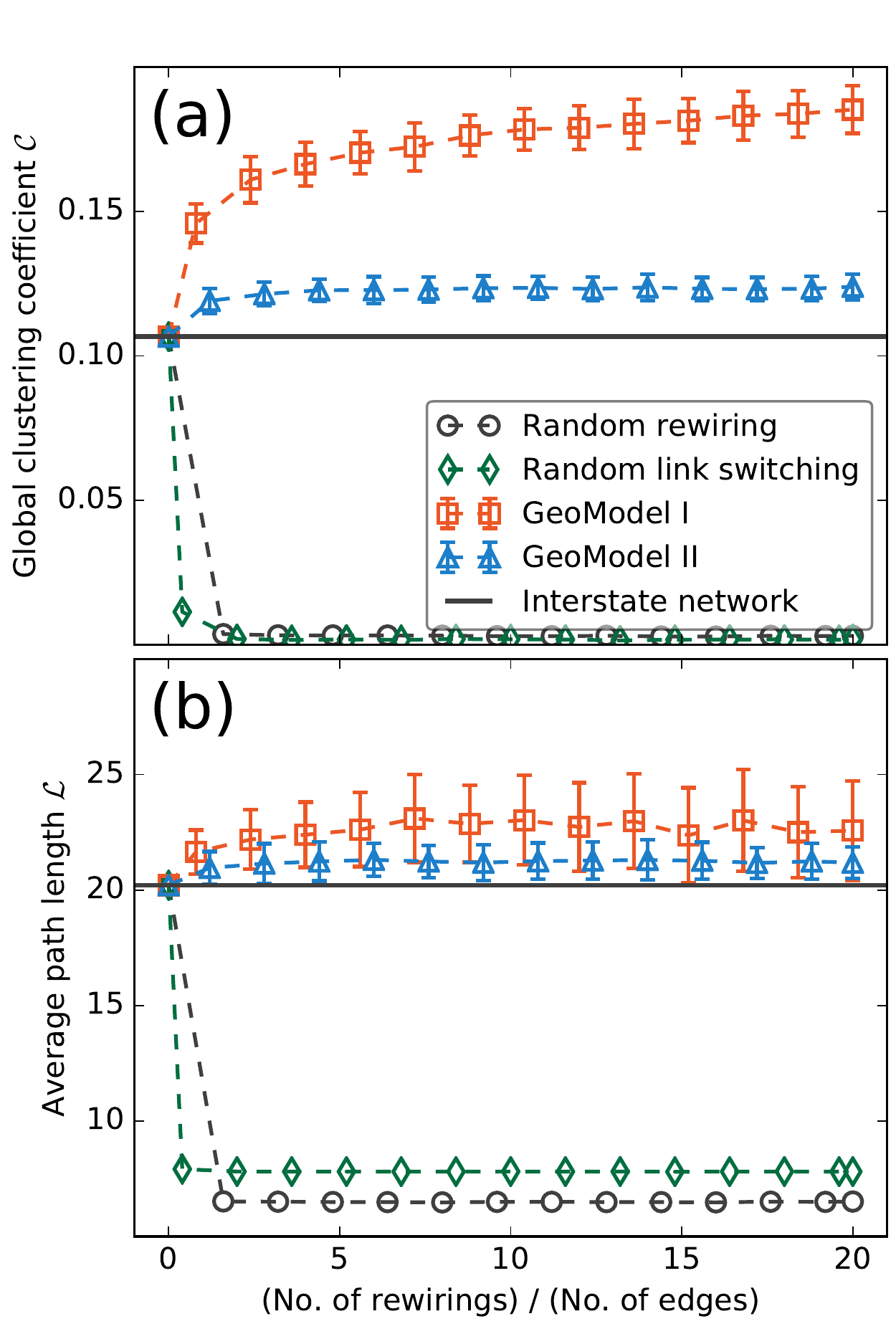}
	\caption{(Color online) Evolution of (a) global clustering coefficient $\C$ and (b) average
		path length $\L$ with the number of rewirings averaged over ensembles of $n=100$
		surrogates generated from the interstate network by applying the different
		random network models (dashed lines). For GeoModel I and GeoModel II the tolerances are set to
		$\epsone=0.17$ and $\epstwo=0.24$, respectively.
		Scatter symbols denote the mean value of 
  each measure. Error bars indicate one standard deviation and are shown if
	their size exceeds that of the symbol.
	Solid lines indicate the value of $\C$ and $\L$ in the
	original network. \label{fig:fig_04}}
\end{figure}

With the two tolerances $\epsone=0.17$ and $\epstwo=0.24$ estimated for
applying GeoModel I
and GeoModel II to the interstate network, we now investigate the evolution of $\C$
and $\L$ with an increasing number of rewiring steps for the four different random
network models (Fig.~\ref{fig:fig_04}). 
Generally, we note that random rewiring and random link switching converge 
towards a state where there is hardly any further fluctuation in the evolution of $\C$
and $\L$ after less than $2M$ steps of rewiring (Fig.~\ref{fig:fig_04}).
Similarly, GeoModel II converges after $5M$ steps. Only for GeoModel I, we note small
fluctuations in the average evolution of $\L$ (Fig.~\ref{fig:fig_04}(b)) and a
slow saturation in the average evolution of $\C$ (Fig.~\ref{fig:fig_04}(a)) up to the
maximum value of $r=20M$ steps of rewiring.

We note that surrogate networks obtained
from random rewiring and random link switching do not capture well the
macroscopic characteristics of the interstate network indicated by large
deviations of $\C$ and $\L$ from their original values
(Fig.~\ref{fig:fig_04}(a),(b)). In fact, with respect
to the global clustering coefficient $\C$, the two models perform equally 
badly (Fig.~\ref{fig:fig_04}(a)). For the average path length $\L$, the additional
constraint of a preserved degree sequence when applying random link switching
yields a slight improvement over the process of random rewiring as in average the
surrogate networks' values of $\L$ are closer to that of the original network.

Additionally taking into account spatial constraints on the lengths of links in
the random networks and, hence, applying GeoModel I and GeoModel II yields
macroscopic characteristics of the surrogates that are much closer to those of
the original network (Fig.~\ref{fig:fig_04}). Specifically, GeoModel I already
gives values of $\L$ very close to that of the original interstate network
(Fig.~\ref{fig:fig_04}(b)), while the resulting values of $\C$ still deviate
strongly from their target (Fig.~\ref{fig:fig_04}(a)). The additional constraint
of a preserved local link length distribution $P_i(l)$ overcomes this issue and
GeoModel II provides surrogate networks that, in addition to $\L$, also
approximate $\C$ in good agreement with the original network. However, slight
differences in the two quantities obtained from GeoModel II in comparison with
the original network's characteristics are still present. Additionally
constraining the algorithm to also preserve a network's degree
correlation~\cite{zamora-lopez_reciprocity_2008} might further improve the
agreement between the surrogates and the original network. An investigation of
such higher order effects remains as a subject of future research. We also note
that GeoModel I and GeoModel II tend to give too high the values of $\L$ and
$\C$ at least for the particular case of the interstate network. This effect
could be related to optimization principles, such as the minimization of
intersection crossings for road networks, underlying the original network that
are not accounted for by the surrogate networks' construction mechanism. Future
studies should address in more detail, for what types of networks GeoModel I
and GeoModel II possibly show systematic positive or negative biases with
respect to the target values of, e.g., global clustering coefficient and
average path length.

\subsection{Airline network}\label{sec:airline}
\begin{figure}[t]
	\includegraphics[width=.8\linewidth]{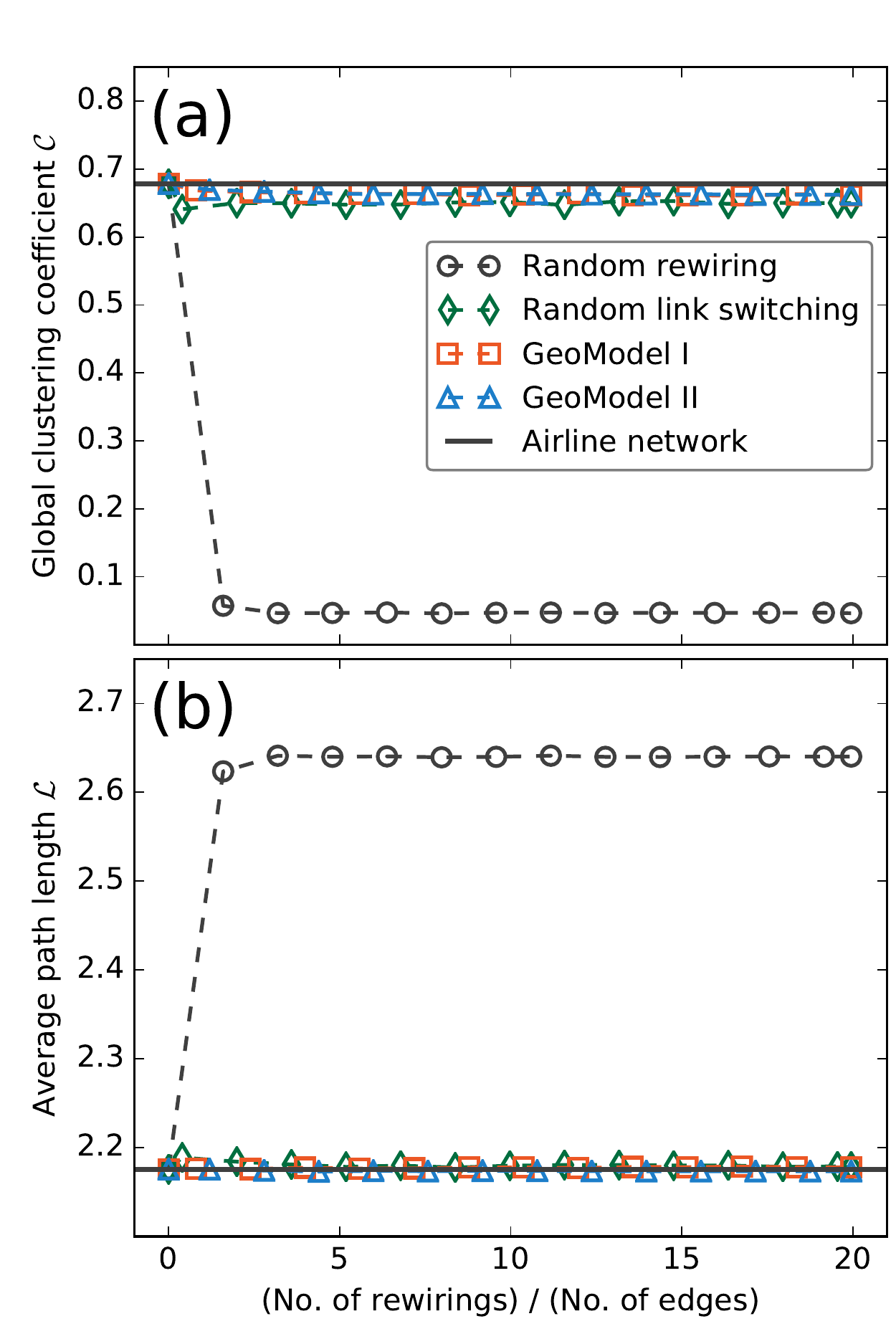}
  \caption{(Color online) Same as Fig.~\ref{fig:fig_04} but for the airline
	network and tolerances of $\epsone=0.04$ and $\epstwo=0.06$. Error bars are
	not shown as they do not exceed the size of the symbols.\label{fig:fig_05}}
\end{figure}
We now apply the same procedure as discussed before to the airline network and
compute the evolution of global clustering coefficient $\C$ and average path
length $\L$ with an increasing number of rewiring steps for the four different
random network models (Fig.~\ref{fig:fig_05}).  We note a fast convergence
towards a state with hardly any further fluctuations in the average evolution
of $\C$ and
$\L$ for all four random network models.  As for the interstate network, we
find that random rewiring does not produce surrogate networks that capture
well the macroscopic characteristics of the airline network. However, in
contrast to the former case, random link switching already reproduces very well
both macroscopic quantities $\C$ and $\L$. GeoModel I and GeoModel II slightly
improve these results further, but for the present case of the airline network a
prescribed degree sequence already produces surrogate networks with properties
close to those of the original network. Thus, for the airline network the
spatial embedding of the nodes and the resulting characteristic distribution of
link lengths is less important for its macroscopic properties as compared
to the US interstate network.

\subsection{Inter-comparison between different spatial networks}
\begin{figure}[t]
	\includegraphics[width=\linewidth]{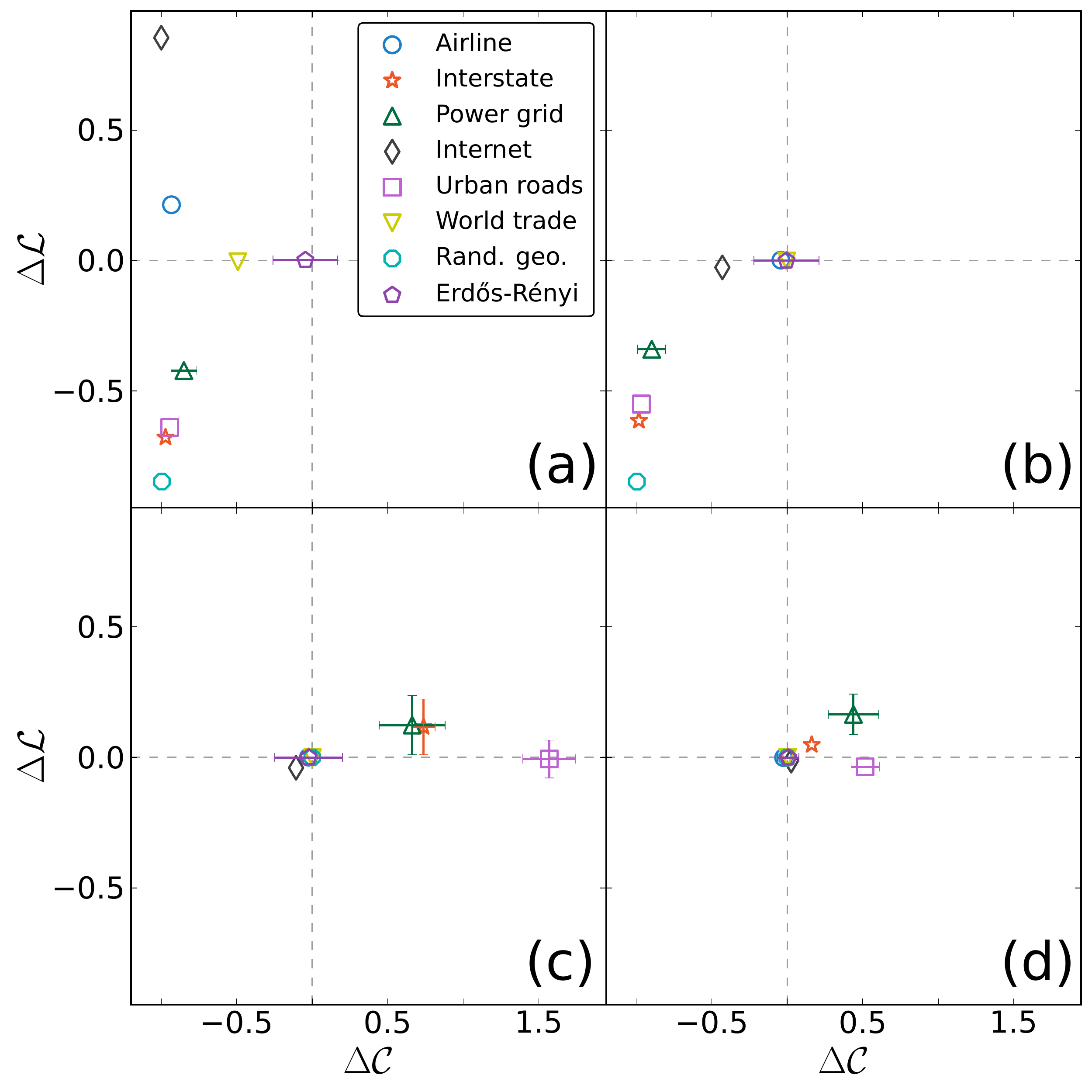}
	\caption{(Color online) Average relative deviation of global clustering
		coefficient $\Delta\C$
		and average path length $\Delta\L$ from the respective original values
		computed over an ensemble of $n=100$ surrogate networks after $20M$
		successful rewirings by applying (a) random rewiring, (b) random link
		switching, (c) GeoModel I and (d) GeoModel II.  The tolerances $\epsone$ and
		$\epstwo$ used for each network and random network model are given in
		Tab.~\ref{tab:tab_01}. Error bars denote the standard deviation in
		$\Delta\C$ and $\Delta\L$ and are shown if their size exceeds that of the
		corresponding symbol.
		}\label{fig:fig_06}
\end{figure}
As in the previous sections, we now compute for each of the networks
under study (Tab.~\ref{tab:tab_01}) the evolution of global clustering
coefficient $\C$ and
average path length $\L$ by evaluating ensembles of $n=100$ realizations of
each network model and using $r=20M$ rewiring steps.  To give a comprehensive
summary, we compute for each network and network model the average relative
deviation $\Delta\C$ and $\Delta\L$ from the respective original network's
values,
\begin{align}
	\Delta\C &= \frac{\langle\C_{sur}\rangle - \C_{orig}}{\C_{orig}}\\
	\Delta\L &= \frac{\langle\L_{sur}\rangle - \L_{orig}}{\L_{orig}}.
\end{align}
Figure~\ref{fig:fig_06} summarizes the results for all spatial networks under
study. As expected, the \ER network's topological features are already well reproduced by
random rewiring, while this is not the case for all other networks
(Fig.~\ref{fig:fig_06}(a)). In all cases, the global
clustering coefficients $\C$ of the surrogate networks are lower than those of the
respective original networks (indicated by negative values of $\Delta\C$ in
Fig.~\ref{fig:fig_06}(a)), which is in accordance with the fact that networks generated from random rewiring are expected
to display a clustering coefficient close to their link
density~\cite{albert_statistical_2002}. Remarkably, the
average path length of the world trade network is also already well
reproduced by random rewiring (resulting in $\Delta\L$ close to zero in
Fig.~\ref{fig:fig_06}(a)), which is likely due to its large link density of
$\rho\approx0.4$. We note that for the Internet and the airline network, the
randomly rewired surrogates have a positive bias for the average path length
$\L$, while for
all remaining networks, a negative bias is found. 

As discussed in Section~\ref{sec:airline}, the process of random link switching
reproduces well the macroscopic properties of the airline network
(Fig.~\ref{fig:fig_06}(b)). The same observation also holds for the world trade
and, as expected, for the \ER network.  Additionally, the average path length
of the Internet is already well captured by random link switching, too. Thus,
the topological features of these networks are already well expressed in terms
of their degree distribution and the spatial embedding of the nodes seems to
have little influence on the average path length $\L$ and the global clustering
coefficient $\C$. For the four other networks (power grid, urban roads,
interstate, and random geometric graph), only slight improvements are visible
when comparing the relative deviations $\Delta\C$ and $\Delta\L$ obtained by
applying random rewiring with those for random link switching (compare
Fig.~\ref{fig:fig_06}(a) and Fig.~\ref{fig:fig_06}(b)).

Additionally taking the effects of the nodes' spatial embedding into account,
we find that GeoModel I generates random surrogates of all networks under study
for which the average path length $\L$ already becomes very close to its
original values (Fig.~\ref{fig:fig_06}(c)). However, while	for the airline,
Internet, world trade and \ER networks, the surrogates are also in good
agreement with respect to deviations in the global clustering coefficient, we
observe that GeoModel I still shows a positive bias of $\C$ for the
remaining networks. Thus, for the aforementioned networks, the global link
length distribution $P(l)$ already determines the expected value of the average
path length $\L$, while the global clustering coefficient $\C$ is not yet
explained sufficiently.

This mismatch is, however, to a large extent addressed by the usage of GeoModel
II (Fig.~\ref{fig:fig_06}(d)). We now find for all networks a reduction of the
deviation in $\C$ as compared with the application of GeoModel I (compare
Fig.~\ref{fig:fig_06}(c) and Fig.~\ref{fig:fig_06}(d)). This means that ultimately,
in addition to the global link length distribution $P(l)$, the local link
length distribution $P_i(l)$ predetermines in most cases and to a large extent
the value of the global clustering coefficient $\C$.

In summary, we have identified a class of spatial networks (including the
spatially embedded Erd\H{o}s-R\'enyi, airline, world trade and Internet
network) for which random network models that
do not take into account any spatial embedding of the nodes already generate
surrogates with global clustering coefficients and average path lengths similar
to those of the original networks. For a second class of networks (the power
grid and random geometric graph as well as the interstate and urban road
network) only taking the spatial structure of the original network explicitly
into account in terms of GeoModel I and/or GeoModel II produces surrogates with
global clustering coefficients and average path lengths comparable with those
of the respective original networks.  Remarkably, we find that GeoModel I
serves to reproduce well the average path lengths of the aforementioned
networks, while only the application of GeoModel II produces network surrogates
for which the global clustering coefficient becomes also close to the
respective original network's value.

We emphasize that the first class of networks, which includes the airline
network, is generally non-planar. In contrast, those networks where spatial
embedding is found to have a larger influence on macroscopic properties are
almost or even fully planar. This hints to a probable relationship between the
macroscopic properties studied in this work and the planarity of networks. An
extension of the proposed models to also address these effects would help to
further comprehend the mechanisms generating the observed network topologies.
Note again that the models presented in this study are designed such that
constraints are added to lower-order models in a consecutive manner (GeoModel I
adds one additional constraint $\textbf C_2$ to random link switching while
GeoModel II adds one constraint $\textbf C_3$ to GeoModel I). This nested model
structure allows for a
precise tracking of the effects that cause the surrogates' global network
characteristics to converge into those of the original network. However, it is
not feasible to address the planarity constraint in a similar fashion. This
constraint rather requests to also consider the actual geographical position of
each node in the metric space which is not dealt with so far. In addition,
further constraints would continuously reduce the set of possible surrogates
that can be generated from a given network to an even lower size than
GeoModel II does. Thus, in order to also address the issue of planarity a
different approach of rewiring and network generation than the one presented
here seems to be more appropriate. Future work should thus focus on this
important issue and evaluate possible construction mechanisms for surrogate
networks that preserve the planarity of a given network in addition to the
spatial constraints discussed in this work.

\subsection{Further evaluation of link length distributions}
\begin{figure}[t]
	\includegraphics[width=.75\linewidth]{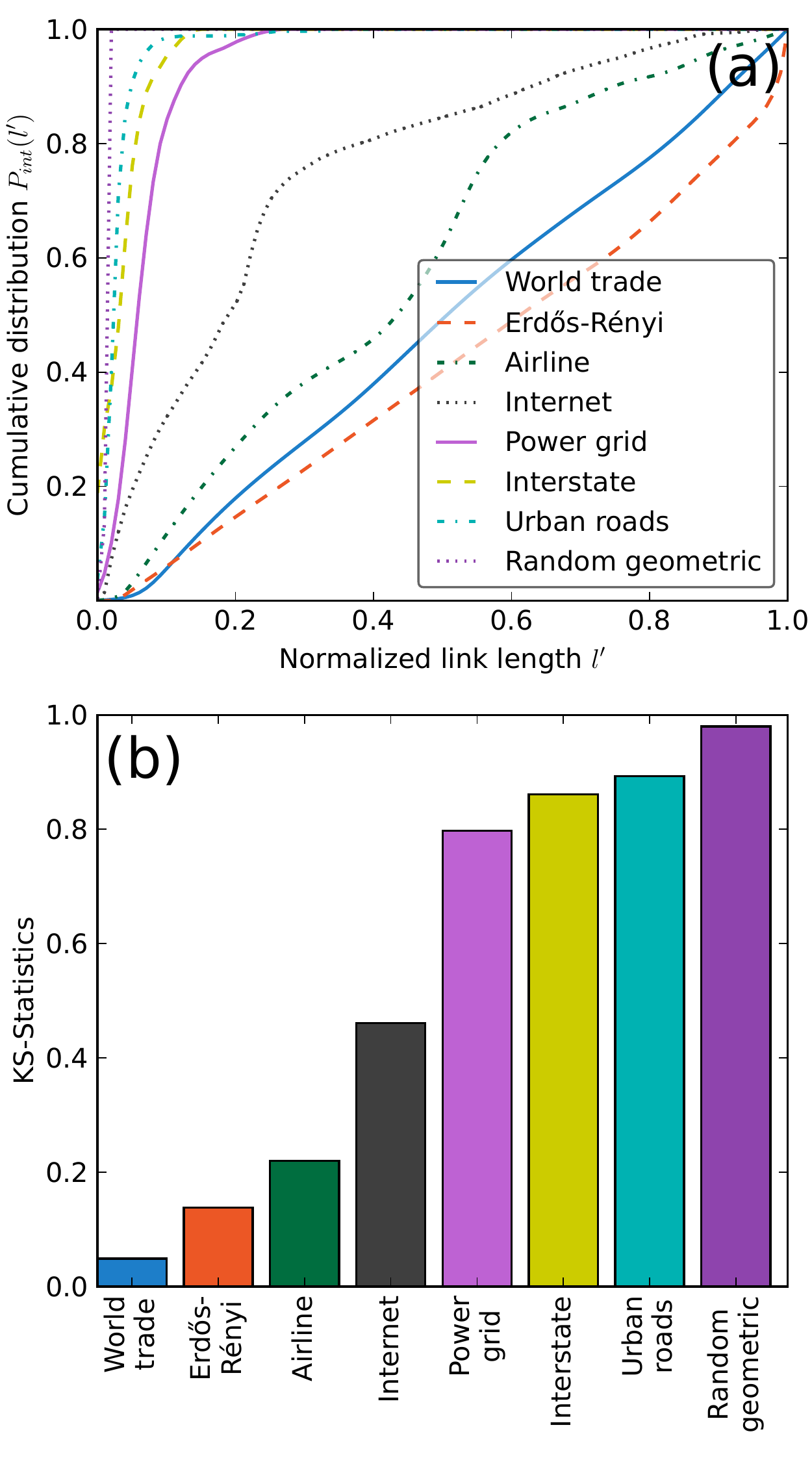}
	\caption{(Color online) (a) Cumulative intrinsic probability distribution
		$P_{int}(l')$ for the different networks under study. (b) KS
		statistics indicating the maximum distance between each cumulative
		probability distribution $P_{int}(l')$ and that of a random uniform distribution for each
		network under study (sorted in ascending order). }\label{fig:fig_07}
\end{figure}
All results obtained in the previous sections indicate a classification of
the networks under study into two different classes -- (i) those where spatial
embedding largely determines the networks' topology and (ii) those which are less
affected by spatial constraints. We now aim to further disentangle this
typology based on the original networks' link length distributions.
For this purpose, we first normalize for each network the distances $d_{ij}$ between
all nodes $i$ and $j$ by the maximum possible distance $\max\{d_{ij}\}_{i,j}$
and thus obtain normalized distances $d_{ij}'\in [0,1]$. We then estimate the
distribution $p(l')$ of normalized link-lengths $l'\in [0,1]$ for
each network using a Gaussian kernel with a bandwidth chosen
according to Scott's Rule~\cite{scott_multivariate_2015}. 

Let us now assume $p(l')$ to factorize into two disjoint probability
distributions: First, $p_{geo}(l')$ indicates the probability to find a
distance in the interval $[l', l'+dl']$ between two randomly chosen nodes $i$
and $j$ that are not necessarily connected by a link, i.e., $p_{geo}(l')$
denotes the geographic inter-node distance distribution between all possible pairs of
nodes in the network. Second, let $p_{int}(l')$ denote the \textit{intrinsic}
probability of the network to form a link with a length within the interval
$[l', l'+dl']$ between two nodes $i$ and $j$ that display a corresponding
geographical distance taken from the same interval. 

The two probability densities $p(l')$ and $p_{geo}(l')$
can be easily estimated from the data. Recall that $p(l')$ denotes the distribution of
all link lengths, while $p_{geo}(l')$ give the distribution of all
inter-node distances. The quantity of interest in the scope of this
work is the residual probability $p_{int}(l')$ which remains after factorizing
out the general spatial constraints of the systems. From
\begin{align}
p(l')=p_{geo}(l')p_{int}(l')
\end{align}
we immediately compute $p_{int}(l')$. The corresponding cumulative distribution
functions $P_{int}(l')$ for all networks under study are displayed in
Fig.~\ref{fig:fig_07}(a). 

We note that the four networks (power grid, interstate, urban roads and the
random geometric graph) that were previously identified as being strongly
influenced by the spatial embedding of the system display a strong increase in
$P_{int}(l')$ for low values of $l'$. This signature is typical for exponential
distributions and, thus, the intrinsic probability for links to be present in
the network depends strongly on the inter-node distances even after accounting
for all trivial geometric factors
(Fig.~\ref{fig:fig_07}(a)). In contrast, the world trade, airline and Erd\H{o}s-R\'enyi network display a cumulative
intrinsic link-length distribution $P_{int}(l')$ that increases almost
linearly, a typical signature of a uniform distribution. Thus, for
these networks, the actual linking probability that emerges after eliminating
all trivial spatial effects is mainly random and does not depend much on the
spatial embedding of the system (Fig.~\ref{fig:fig_07}(a)). 

The Internet presents a case slightly in between these two qualitatively
different behaviors, exhibiting neither a strictly linear increase in
$P_{int}(l')$ nor a sharp increase for
small values of $l'$. This aligns well with the observation made from
Fig.~\ref{fig:fig_06}(b), where the Internet sticks out in the sense that its average
path length is already fully determined by the system's spatial embedding while
its global clustering coefficient is not. This signature is unique to the
Internet when compared to all networks under study. 

A summary of these findings is shown in Fig.~\ref{fig:fig_07}(b) which displays
the KS distances for each of the intrinsic cumulative distribution functions
$P_{int}(l')$ from a uniform distribution. We
clearly note the low values for the world trade, airline and Erd\H{o}s-R\'enyi
network, the intermediate value for the Internet and large values, i.e., large
deviations from purely random behavior for the power grid, interstate, urban roads and the
random geometric graph. Thus, the observed of link length distributions
underline our findings from the previous sections and point again to a distinction
of spatial networks into different distinct classes.   

From a mechanistic point of view, it thus seems that there exists one class of
networks for which global network characteristics can already be well
reproduced by putting links between nodes according to a probability
distribution that is purely determined by their distances. However, for a
second class a second construction principle is superimposed that modulates
this node-distance distribution such that short links are preferred over
long-range connections.  

We are well aware of the fact that the assumption of independence between the
intrinsic and geographic probability distributions $p_{int}(l')$ and
$p_{geo}(l')$ is very strong and needs proof in the course of upcoming
research. Specifically, this factorization assumes that nodes and links in a
spatially embedded network are in a sense created independently from each
other. This might to some extent be true for, e.g., the world trade network
where first countries emerge and trade connections are then put between them,
or the airline network, where already existing cities are in a second step
connected by flights. In contrast, for systems such as road networks the
creation of links representing streets and nodes representing intersections is
naturally closely entangled and thus the above assumption certainly needs
further evaluation. However, the findings from this section are mainly meant to
present a first approximation of possible mechanisms behind the observed
classification of spatial networks into different groups and suggest an
interesting direction to further investigate the nature of spatially embedded
networks. Future work should pick up these lines of thought to investigate more
thoroughly the mechanisms that drive the emergence of links in those different
types of networks. 

\section{Conclusion}\label{sec:conclusion}
We have introduced two novel models to generate random surrogates of a given
spatial network that preserve either the global or the local distribution of
link lengths between individual nodes and, hence, explicitly take into account
the embedding of the network in some metric space. We have characterized the
macroscopic properties of the resulting surrogates by means of the global
clustering coefficient and the average path length and compared these values to
those of the original networks from which the surrogates were constructed. For
reference, we have utilized iterative random rewiring and random link switching
to produce random networks similar to \ER networks and the configuration model,
respectively. 

We have found that for a certain class of spatial networks random link
switching already produces surrogates of comparable macroscopic structure as
the original network. Thus, for these networks the spatial embedding of the
nodes and links is not crucial for explaining their corresponding macroscopic
properties. In contrast, we have identified another class of networks for which
global clustering coefficients and average path lengths are only well
reproduced when applying the newly introduced GeoModel I and/or GeoModel II that
explicitly account for the spatial embedding of the nodes. Hence, for these
networks information about their geometric properties is needed to
sufficiently explain their macroscopic structure. For the latter class of
networks, we have found that their average path length can already
be well reproduced by GeoModel I, while only using GeoModel II enables us to also capture
the global clustering coefficient to a large extent. Our findings align well
with recent studies on the effect of the networks' spatial embedding on the
small-world property of a system~\cite{bialonski_brain_2010}. We confirmed that
the two quantities that are commonly assessed when determining whether a
network displays the small-world property are in many cases to a large extent
already predetermined by the spatial distances between its nodes.

In summary, the surrogate network models  introduced in this work provide an
important step in assessing whether and to which extent global characteristics
of a complex network are already predetermined by statistics associated with
the spatial embedding of its nodes and links.  For future work, it will be of
great importance to study in more detail which classes of networks are
explicitly affected by the nodes' spatial embedding and which are already
sufficiently quantified by some structural quantities such as the degree
distribution. This issue might also imply to specifically impose constraints on
the planarity of the surrogate networks corresponding to the original networks'
topologies. 

We further observed that the optimal tolerance parameter $\epsilon$
(the only parameter of the models we introduced here) varies strongly depending
on the specific networks under study. An assessment of the interplay between
the networks' known topological properties and the estimated values of
$\epsilon$ could help to directly estimate an optimal tolerance circumventing
the need for evaluating KS statistics as
applied in the present work. Additionally, it is of interest to extend the
models presented in this work to also conserve degree
correlations~\cite{zamora-lopez_reciprocity_2008} and to be also applicable to
weighted networks, such as airline networks, where the weight of each link
scales with the number of passengers on the corresponding
connection~\cite{barrat_architecture_2004}. In order to compare different types
of networks, a thorough, concise, and general definition of link weights must
be found as otherwise no useful intercomparison of obtained weighted network
characteristics is possible. In order to develop a corresponding framework,
our models could be combined with existing models for non-spatially embedded
weighted networks~\cite{ansmann_constrained_2011} that follow a similar
strategy of constrained rewiring of a given network as the models presented in
this work.

\begin{acknowledgments}
	MW and RVD have been supported by the German Federal Ministry for 
	Education and Research (BMBF) via the Young Investigators Group CoSy-CC$^2$
	(grant no.~01LN1306A). JFD thanks the Stordalen Foundation and BMBF (project GLUES) for
	financial support. JK acknowledges the IRTG 1740 funded by DFG and FAPESP\@.
	Michael Gastner is acknowledged for providing his data on the airline, interstate,
	and Internet network. Peter Menck thankfully provided his data on the
	Scandinavian power grid. We thank Sven Willner on behalf of the entire \texttt{zeean} team 
	for providing the data on the world trade network. All computations have been
	performed using the Python package
	\texttt{pyunicorn}~\citep{donges_unified_2015} available at
	\url{https://github.com/pik-copan/pyunicorn}.
\end{acknowledgments}

%
\appendix*
\section{Kolmogorov-Smirnoff Test}\label{sec:appendix}
Given two cumulative distribution functions of link lengths $P(l)$ and
$P'(l)$ the Kolmogorov-Smirnoff (KS) statistic $\kappa$ is given as 
\begin{align}
	\kappa =\max_{0 < l < \infty} |P(l) - P'(l)|.
\end{align}
The two distributions are equal at a confidence level $\alpha$
if~\cite{press_numerical_1996}
\begin{align}
	Q_{KS}([M_e + 0.12 + 0.11 / M_e]\kappa) > \alpha. \label{eqn:k_crit}
\end{align}
Here, $M_e = M/2$ is the effective number of links constituting each distribution and
$Q_{KS}$ is given as,
\begin{align}
	Q_{KS}(x)=2\sum_{j=1}^m (-1)^{j-1}\exp(-2j^2x^2).
\end{align}
In theory, the above sum has infinitely many entries, $m=\infty$. In this work
we set $m=100$ to obtain an acceptable approximation.
\end{document}